\documentclass[prb,twocolumn,preprintnumbers,amsmath,amssymb]{revtex4}
\usepackage{graphicx}
\usepackage{color}
\usepackage[sort&compress]{natbib}

\newcommand{\meV}{\ensuremath{\,\mbox{meV}}}
\newcommand{\neV}{\ensuremath{\,\mbox{neV}}}
\newcommand{\mueV}{\ensuremath{\,\mu\mbox{eV}}}
\newcommand{\mT}{\ensuremath{\,\mbox{mT}}}
\newcommand{\nm}{\ensuremath{\,\mbox{nm}}}

\newcommand{\ns}{\ensuremath{\,\mbox{ns}}}

\newcommand{\vc}[1]{\ensuremath{\vec{#1}}}

\newcommand{\Hop}{\ensuremath{\mathcal{H}}}
\newcommand{\Hmb}{\ensuremath{\mathcal{H}}}

\newcommand{\Pth}{\ensuremath{P_{th}}}
\newcommand{\depsTarget}{\ensuremath{\Delta\epsilon_{\mbox{\scriptsize target}}}}

\begin{document}

\title{Implications of Simultaneous Requirements for Low Noise Exchange Gates in Double Quantum Dots}
\author{Erik Nielsen, Ralph W.~Young, Richard P.~Muller and M.~S.~Carroll}
\affiliation{Sandia National Laboratories, Albuquerque, New Mexico 87185 USA}
\date{\today}

\begin{abstract}
Achieving low-error, exchange-interaction operations in quantum dots for quantum computing imposes simultaneous requirements on the exchange energy's dependence on applied voltages.  A double quantum dot (DQD) qubit, approximated with a quadratic potential, is solved using a full configuration interaction method.  This method is more accurate than Heitler-London and Hund-Mulliken approaches and captures new and significant qualitative behavior.  We show that multiple regimes can be found in which the exchange energy's dependence on the bias voltage between the dots is compatible with current quantum error correction codes and state-of-the-art electronics. Identifying such regimes may prove valuable for the construction and operation of quantum gates that are robust to charge fluctuations, particularly in the case of dynamically corrected gates.
%  The exchange energy's dependence on dot size and spacing is also investigated, and approaches to reduce sensitivity to these quantities are discussed.
%    Implications to a dynamically decoupled $z$-rotation gate is discussed.
%(1) the exchange energy $J$ is relatively insensitive to fluctuations in the bias voltage between the dots for positive and negative values of $J$ simultaneously, and (2) the magnitude of $J$ corresponds to qubit rotation times that are much greater than electronics jitter.  
\end{abstract}

%\pacs{todo}
\maketitle

%INTRODUCTION
\section{Introduction\label{secIntro}}
A quantum bit (qubit) typically encodes information in a two-level system.  The exchange energy between quantum dots was first suggested as sufficient to perform a universal gate set (unitary coherent manipulations of one and two qubits for logical operations) by Levy,\cite{JLevyUniversal_2002} and subsequent exchange-based proposals for solid-state architectures have been suggested by Loss-DiVincenzo,\cite{BurkardLossDivincenzo_1999} Kane,\cite{KaneNature_1998} and Taylor.\cite{TaylorNatPhys_2005}  The exchange interaction causes a splitting between quantum states called the \emph{exchange energy}, which we denote $J$. Qubit rotations are performed experimentally by electrically increasing the exchange energy for short times.\cite{PettaScience_2005}

Several important noise sources that can produce error in the exchange operation include charge fluctuations (\emph{e.g.}~random telegraph noise, Johnson and shot noise), inaccuracy in electronics control (\emph{e.g.}~ringing and over/undershoot) and rotations due to inhomogeneous fields.

%Error correction schemes
Quantum error correction (QEC) schemes have been developed to cope with noise and errors in future quantum circuits.\cite{NielsenChaung} A quantum error correction code introduces redundancy in the qubit information providing the ability to correct for errors through majority vote checks on the redundant basis bits.\cite{Knill_1997}  These coding schemes are believed to be a necessary component of any future quantum computer because of the fragile nature of qubits. However, the codes provide benefit only for cases when the qubit gate error rate is less than a threshold value $\Pth$ above which the error correction circuit is more faulty than a bare qubit.   Thresholds have been estimated for a number of cases and almost ubiquitously predict very strict limits on the tolerable error in the gate operations (\emph{e.g.}~$\Pth=2\times 10^{-4}$ and $\Pth=2\times 10^{-5}$ from Refs.~\onlinecite{TaylorNatPhys_2005},\onlinecite{LevyClassicalConstraints_2009}).

A number of approaches are being pursued to minimize errors in qubit gates to achieve operations that are sufficient to realize the benefits of quantum error correction codes.  The exchange gate couples the charge degree of freedom to the spin degree of freedom, which is useful for electronic control of the spins but also exposes the gate to errors induced by the electrodynamics of the system.  A number of strategies have been proposed in the literature to address different forms of errors (\emph{e.g.}~large $J$ for fast rotations relative to noise sources,\cite{JLevyUniversal_2002} $dJ/d(\mbox{bias}) \approx 0$ to suppress the impact of voltage fluctuations similar to those due to detuning\cite{StopaImmunity_2008}, and multiple rotation velocities for dynamically corrected gating\cite{KhodjastehViola_2009}).  These strategies introduce a number of constraints, and it is not obvious that all of them can be simultaneously implemented in a double quantum dot given the physics of the system.  In this paper we show that the simultaneous constraints are consistent with a semi-qualitative model of a double quantum dot system.   
%- revise this?

A configuration interaction (CI) method is used to study $J$ as a function of parameters which specify a double quantum dot system.  This CI method is more general than Heitler London (HL), Hund Mulliken (HM), and Hubbard model approaches, and is found invaluable to accurately calculate, in the single-valley case, energies for the bias range approaching and within the regime where there is two-electron occupation of one dot.  The two-electron occupancy regime is relatively insensitive to the inter-dot bias ($dJ/d(\mbox{bias}) \approx 0$), making it an important regime to accurately calculate.  Furthermore, this method is less computationally demanding than techniques requiring a large mesh, allowing a tractable search for robust exchange interaction parameters in the double dot system.   

We begin by describing our DQD model in section \ref{secModel}, and outlining the CI method used to solve it in section \ref{secCalc}.  We then develop in section \ref{secConstraints} the constraints placed on a DQD's exchange energy by quantum error correction codes and controlling electronics.  Results are presented in section \ref{secResults}, and analyzed using the noise constraints. Finally, we discuss implications and a complementary approach to noise mitigation in section \ref{secDiscussion}, and end with summary and conclusions in section \ref{secConclusion}.

%MODEL
\section{Model\label{secModel}}
A lateral quantum dot singlet-triplet qubit qualitatively similar to that described by Taylor et al.\cite{TaylorNatPhys_2005} is examined in this paper.  To provide a semi-quantitative analysis we use gallium arsenide material constants $m^* = 0.067\,m_e$ and $\kappa=12.9$.  The computational basis (\emph{i.e.}~the levels of the effective 2-state system) consists of the two-electron singlet and $S_z=0$ triplet states of lowest energy.  $J$ is the splitting between these two states.  Note that the triplet states with $S_z=\pm 1$ are split off with a dc magnetic field (typically of order $100\mT$).  The qubit's effective many-body Hamiltonian is given by
\begin{equation}
%Many-particle Hamiltonian
\Hmb = \Hop_1 + \Hop_2 + \frac{e^2}{\kappa r_{12}} \,\, ;\,\, \Hop_i = \frac{\vc{P}^2}{2m^*} + V(\vc{r})+ \frac{e}{m^*}\vc{S} \cdot \vc{B}
\label{eqMBHam}
\end{equation}
where $r_{12}=|\vc{r}_1-\vc{r}_2|$ and $\kappa$ is the GaAs dielectric constant.  In $\Hop_i$, $\vc{P}=\vc{p}-e\vc{A}$, and $\vc{p} = (p_x,p_y)$ and $\vc{r} = (x,y)$ are the usual momentum and position operators of the $i^{\mathrm{th}}$ electron. $\vec{A}$ is a vector potential for the magnetic field $\vc{B} = \vc{\nabla} \times \vc{A}$, and $V$ is the electrostatic potential.  A constant perpendicular field $\vc{B}=B\hat{z}$ is considered here, and we restrict ourselves to two dimensions. %More??
%, although the addition of a gradient in $\vc{B}$, important for certain qubit operations (\emph{e.g.}~$x$-rotation), will be analyzed in later work.

\begin{figure}[h]
\begin{center}
\includegraphics[width=2.2in,angle=270]{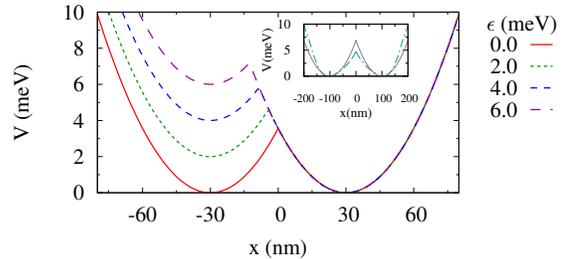}
\caption{DQD potential along x-axis, $V(x,0)$, for $L=30\nm$ and $E_0=3.0\meV$. The inset shows the potential obtained by solving Poisson's equation for both electrons and holes using a commercially available solver for an accurate DQD placement of gates and insulators.  The dotted line is a double-parabolic potential and shows that a potential of this form is a relatively good approximation to the potentials expected in real devices. \label{figPotential}}
\end{center}
\end{figure}

The electrostatic potential $V$ is generated by lithographically formed gates near the semiconductor interface.  By applying different voltages to these gates at different times, the shape of $V$ and the exchange energy can be tuned to perform operations on the qubit (\emph{e.g.} see Ref.~\onlinecite{PettaScience_2005}).  We idealize $V$ as the minimum of two parabolic dots,
\begin{equation}
V(x,y) = \frac{1}{2}m^*\omega^2 \left[ \min\left( (x-L)^2+\epsilon, (x+L)^2 \right) + y^2\right]\,.
\end{equation}
The parameters $\epsilon$, $L$, and $\omega$, correspond to the bias, inter-dot distance, and frequency of the ground state as well as a measure of the confining well potential, respectively. Cuts of the 2D potential along the $x$-axis for different $\epsilon$ are shown in Fig.~\ref{figPotential}.  One reason for choosing this parametrization is so that all three of these aspects can be varied independently. %This is one of the simplest forms for $V$ which has the general features of a realistic DQD potential. 
We also define the dot size $d = \sqrt{\hbar^2/(m^*E_0)}$ as the width of the ground state probability distribution in a parabolic well with confinement energy $E_0$, where $E_0=\hbar\omega$.  
%Drop?
%We have in mind a device where a qubit operation is carried out by a sequence of gate voltages which modulate the bias between the dots while keeping their separation and shape relatively constant.  Thus, our focus is on the behavior of $J$ as $\epsilon$ varies and $L$, $E_0$, and $B$ are fixed.

%CALCULATION (CI DESCRIPTION)
\section{Calculation Method\label{secCalc}}
\begin{figure}[h]
\begin{center}
\includegraphics[width=2in,angle=270]{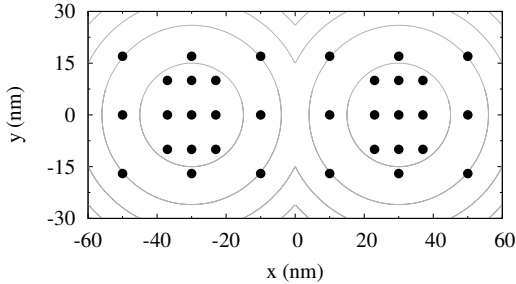}
\caption{The relative positions of the Gaussian basis elements used in our CI method when $n_G=36$. The center of each Gaussian is marked by a solid circle, and equipotential lines show the dot locations.  The 18 elements on each dot are arranged in two concentric $3\times 3$ grids which can have different spacings in the $x$- and $y$-direction, as shown (there are two points at the center of each dot).  The grid spacings are chosen to minimize the lowest singlet or $S_z=0$ triplet energy, as described in the text.\label{figBasisConfig}}
\end{center}
\end{figure}

The CI method used to solve the system Hamiltonian \eqref{eqMBHam} can be decomposed into several steps.  The initial and most complicated step is constructing a basis of $n_G$ s-type Gaussian functions.  Each basis element $g$ is parametrized by a position $(x_g,y_g)$ and exponential decay coefficient $\alpha_g$
\begin{equation}
g(x,y) = N e^{-\alpha_g(x-x_g)^2}e^{-\alpha_g(y-y_g)^2} e^{\frac{ieB}{2\hbar}\left(y_0x-x_0y\right)}
\end{equation}
where $N$ is a normalization coefficient and $B$ is the magnetic field.  A 2D mesh of points (given as input) specifies the positions of the elements in terms of two length scales, $a_x$ and $a_y$, which can be thought of as effective lattice constants of the mesh.  There is a mesh point at each of the dot centers, and we denote the exponential coefficient of the first basis element located at each of these points as $\alpha_c$.  The exponential coefficient for the first element at all other points is denoted $\alpha_o$ (generally $\alpha_o \ne \alpha_c$).  When there are multiple basis elements at a point, the exponential coefficients of additional elements are found by multiplying the previous element's coefficient by a constant factor $\lambda$.  Together the parameters $a_x$, $a_y$, $\alpha_c$, $\alpha_o$, and $\lambda$ specify a Gaussian basis which is used in subsequent steps.  The final values of these parameters are found by optimization of the full many-body energies, as described below.  In all the results presented here $n_G=36$, with 18 elements on each dot arranged in two concentric $3\times 3$ grids (so there are two elements on each dot center), as shown in Fig.~\ref{figBasisConfig}.

%This results in an optimized Gaussian basis to be used in subsequent steps.
%minimizing the energy of the lowest singlet and triplet $S_z=0$ many-body state with respect to all of these variables simultaneously.
%TODO: Insert explicit formula here??

Once a Gaussian basis is chosen, the single-particle Hamiltonian, minus the anomalous Zeeman term, is solved in this basis.  The lowest $n=2n_G$ (including spin degeneracy) of the resulting single-particle states are taken as an orthonormal (single-particle) basis, and all possible 2-particle Slater determinant states are constructed, forming a $n_{MB}$-dimensional two-particle basis, where $n_{MB} = {{n}\choose{2}}$.  Lastly, the full many-body Hamiltonian is diagonalized in this basis.  This method constitutes a full CI with respect to the Gaussian basis.  To improve convergence, the energies for the lowest singlet and $S_z=0$ triplet states are used to iteratively improve the Gaussian basis chosen initially.  The final energies result from a Gaussian basis which is the direct sum of two smaller bases, one which minimizes the energy of the singlet and the other which minimizes the energy of the triplet. Complete details of this procedure will be given elsewhere.\cite{NielsenInPrep}  

Our results are only semi-quantitative, since the exact form of the potential is unknown and the problem is only approximately two-dimensional.  However, the results give a more accurate qualitative and semi-quantitative picture than previous variational approaches, and are sufficient to resolve whether or not regimes exist which are robust to charge noise.  Figure \ref{figJvsEpsB} below compares results of our CI method with Heitler-London and Hund-Mulliken techniques. %TODO: more re: this figure

%NOISE CONSTRAINTS NEEDED FOR ERROR CORRECTION
\section{Error analysis\label{secConstraints}}

\subsection{Exchange (rotation) gate}
%rigorous defn of J ~ 0
One model of an ideal exchange gate operation is to increase $J$ from (near) zero to a finite value $J_1$ for a time $\tau$, and then set $J$ back to zero.  This rotates the qubit about an angle $\theta = J_1\tau/\hbar$, assuming that the exchange energy is dominant and therefore defines the axis of rotation.
%Keep this change?
%Controlling electronics dictate a minimum achievable gate time, $\tau_{min}$, which results in the constraint $|J_1| < J_{max} \equiv \hbar\pi/\tau_{min}$.  A second 

Constraints arise from the error thresholds demanded by error correction codes.  In reality $J$ cannot be perfectly controlled, and for an exchange error $\Delta J$ the rotation angle becomes $\theta+\delta$, where $\delta = \Delta J \tau/\hbar$.  If $J$ is intended to perform a $\theta$-rotation, the gate time $\tau = \hbar\theta/J$, and $\delta$ is of order $\Delta J/J$.

The probability of an error during an exchange gate can be estimated as approximately $\cos\delta \approx \delta^2$.  The error probability should be engineered to be less than the predicted error threshold $\Pth$ of the quantum error correction, which was noted earlier to be dependent on the details of the QEC code and have a wide range of projected values (\emph{e.g.}~$10^{-6}-10^{-4}$).\cite{TaylorNatPhys_2005,LevyClassicalConstraints_2009}  The quantity $\Delta J$ should therefore be targeted to make the ratio $\frac{\Delta J}{J}$ as small as possible such that $\left(\frac{\Delta J}{J}\right)^2$ is at least smaller than the largest QEC threshold, that is, % (\emph{e.g.}~$10^{-4}$)
\begin{equation}
\left| \Delta J/J \right|^2 < \Pth \,. \label{eqNoiseConstraint1}
\end{equation}
The magnitude of $J$ is set by the gate time and target angle of rotation on the Bloch sphere,
\begin{equation}
J = \frac{\hbar\theta}{\tau} \,.  \label{eqJRelation}
\end{equation}
Use of the shortest gate times possible is a common strategy to minimize errors due to time-dependent decoherence mechanisms.  Electronics gate speeds are limited by practical considerations (\emph{e.g.}~jitter and control of rise/fall times) and recent analysis suggests gate times in the range of $1-50\ns$.\cite{PettaScience_2005,Ekanayake10nsPulse_2008,LevyClassicalConstraints_2009}  The exchange energy must therefore be of order $1-0.02\mueV$ for a $\pi/2$ rotation.  

\subsection{No-op (idle) gate}
During a no-op gate, when no rotation is desired, $J$ is ideally zero.  If instead $J$ takes finite value $J_0$, an erroneous rotation $\delta_0 = J_0 \tau_0/\hbar$ will occur over the time period $\tau_0$.  Inserting this into the condition $|\delta_0|^2 < \Pth$ we find that the the magnitude of the exchange must satisfy
\begin{equation}
|J_0|^2 \le \hbar^2\Pth/\tau_0^2\,, \label{eqNoiseConstraintNOOP}
\end{equation} %where $\tau_0$ is the idle gate time
where $\tau_0$ is length of time the qubit is idle.  When $\Pth = 10^{-4}$ this requires $|J_0| \le 6.5-0.13\neV$ for $1-50\ns$ gate times.

\subsection{Simultaneous Constraints}
Since an exchange gate operation involves tuning $J$ between two values (usually zero and a finite value $J_1$), a robust gate requires the existence of at least two \emph{operating points} where Eqs.~\ref{eqNoiseConstraint1} and \ref{eqNoiseConstraintNOOP} are satisfied, respectively, for every needed rotation angle $\theta$.

Additional requirements for dynamically corrected gating (DCG) are also potentially necessary in order to cancel other noise sources such as inhomogeneous quasi-static fields.  At least three $J(\epsilon)$ operating points are desired for DCG, and it is useful, but not necessary, for $J$ to be negative at one of them.\cite{KhodjastehViola_2009,KhodjastehLidarViola_2009}  A critical question theoretically is whether the dependence of $J$ on the quantum dot properties can simultaneously realize several or all of these needs and thereby come closer to fulfilling the strict gate error requirements suggested by present error correction strategies.
%(1) $|J| < 0.2-20\mueV$, and (2) $dJ/d\epsilon < 10^{-8}-10^{-6}$. Condition (1) ensures that the gate rotation times are accurately controllable by state-of-the-art electronics (\emph{e.g.}~accounting for jitter), and (2) is required for robustness to $\epsilon$-noise.

\section{Exchange Energy Results\label{secResults}}
The dependence of $J$ on the electrostatic potential and magnetic field is examined using the CI method to identify whether the predicted DQD exchange energy can meet the anticipated requirements for an exchange gate.  We briefly consider the typical behavior of the exchange energy as a function of system parameters and then we analyze specific noise-robust regimes in detail.

\subsection{Typical behavior}
\begin{figure}[h]
\begin{center}
\includegraphics[width=2.2in,angle=270]{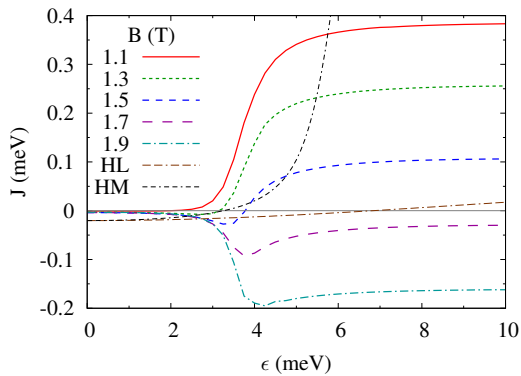}
\caption{The exchange energy $J$ as a function of inter-dot bias $\epsilon$ for $E_0=3\meV$, $L=30\nm$, and values of the magnetic field around the point where the (0,2) singlet and triplet cross. At large and small $\epsilon$ the curve $J$ is relatively insensitive to changes in $\epsilon$.  The transition from positive to negative $J$ in the (0,2) regime (large $\epsilon$) gives rise to a local minimum near $\epsilon=3.5\meV$.  This provides a third region where $J$ is relatively robust to $\epsilon$-variations.  The curves labeled HL and HM show the Heitler London and Hund Mulliken methods for $B=1.1\,\mbox{T}$, respectively. \label{figJvsEpsB}}
\end{center}
\end{figure}

Figure \ref{figJvsEpsB} shows the behavior of $J$ as a function of $\epsilon$ for varying perpendicular magnetic field strengths.  When $B \le 1.1\,\mbox{T}$, the curve has two relatively flat sections at small and large $\epsilon$, where both the singlet and triplet states are in the (1,1) or (0,2) charge sector, respectively.  When the inter-dot barrier is large enough, the quantum states of the DQD will often have an integral number of electrons in each dot.  When there are $n_L$ and $n_R$ electrons in the left and right dots, respectively, the state is said to be in the ($n_L$,$n_R$) charge sector.  Between the flat regions (\emph{e.g.} in the case $B=1.1\,\mbox{T}$) $J$ increases rapidly.  This is because as $\epsilon$ increases the singlet transitions to the (0,2) sector before the triplet, resulting in the DQD potential penalizing the triplet.  The nearly constant value of $J$ at large $\epsilon$ is essentially the exchange energy of a doubly-occupied single dot with confinement energy $E_0$.

Increasing the magnetic field favors the higher angular momentum of the triplet state relative to the singlet.  The magnetic field needed to invert the singlet and triplet levels is lower for (1,1)-states than for (0,2)-states, and this creates a negative-$J$ dip in the $J$ vs.~$\epsilon$ curve at intermediate magnetic field,\cite{StopaImmunity_2008} as shown in Fig.~\ref{figJvsEpsB}.  Since the slope $dJ/dB$ is larger for (0,2)-states, at large enough magnetic field the dip disappears.

\subsection{Noise-robust regimes}
%TODO -- maybe change regime number to roman numeral?
%     -- maybe enumerate regimes since they're referenced so much later?
Given this general dependence of the exchange energy on inter-dot bias, three regimes can be identified which are relatively robust to $\epsilon$-noise:  (I) at low-$\epsilon$, where the electrons are relatively isolated and $J\approx 0$; (II) at high-$\epsilon$, where the singlet and triplet are in the (0,2) charge sector; and (III) at a local minimum, present for certain finite magnetic fields, where the singlet and triplet are between the (1,1) and (0,2) charge sectors.  Whether the $J$ vs.~$\epsilon$ curve in these regions can be made flat enough to meet the stated requirements (Eq.~\ref{eqNoiseConstraint1} or \ref{eqNoiseConstraintNOOP}) is a central question we address in this work.

Let us write the $\epsilon$-dependence of $J$ explicitly, $J=J(\epsilon)$, and define the average change in $J(\epsilon)$ when $\epsilon$ changes by $\Delta\epsilon$ around $\epsilon_0$: $\Delta J(\epsilon_0,\Delta\epsilon) = \frac{1}{2}\left( |J(\epsilon_0+\Delta\epsilon) - J(\epsilon_0)| + |J(\epsilon_0)-J(\epsilon_0-\Delta\epsilon)|\right)$.  Values of $\epsilon$ used as rotation gate operating points must satisfy two criteria.  Using Eq.~\ref{eqJRelation},
\begin{equation}
  |J(\epsilon)| \le J_{max}(\theta,\tau) \equiv \hbar|\theta|/\tau \,, \label{eqMagJConstraint}
\end{equation}
where, as previously, $\theta$ is the gate rotation angle and $\tau$ is gate time.  Secondly, in order to satisfy Eq.~\ref{eqNoiseConstraint1}, $\epsilon$ must also be chosen so that 
\begin{equation}
\Delta\epsilon_{\mbox{\scriptsize ctl}} \le \depsTarget \equiv \max \left\{ \Delta\epsilon : \left|\frac{\Delta J(\epsilon,\Delta\epsilon)}{J(\epsilon)}\right|^2 < \Pth \right\} \label{eqDEpsConstraint}
\end{equation}
where $\Delta\epsilon_{\mbox{\scriptsize ctl}}$ is the bias uncertainty achievable by the controlling electronics.  Given $\tau$, $\theta$, $\Pth$, and $\epsilon$, $\depsTarget$ defines a target value of $\Delta\epsilon$ which must be met in order to ensure noise robustness. In the case of a no-op gate, Eq.~\ref{eqNoiseConstraintNOOP} needs to be satisfied with $J_0$ equal to the exchange energy at and around the operating point, that is,
\begin{equation}
\max \big\{ |J(\epsilon')|^2 : \epsilon' \in [\epsilon+\Delta\epsilon,\epsilon+\Delta\epsilon] \big\} \le \hbar^2 \Pth / \tau_0^2 \,. \label{eqNoiseConstraintNOOPb}
\end{equation}
The quantity $\depsTarget$ can be defined in this case as the maximum value of $\Delta\epsilon$ for which Eq.~\ref{eqNoiseConstraintNOOPb} is satisfied.  Thus, for fixed $\Pth$, $\Delta\epsilon_{\mbox{\scriptsize ctl}}$, $\theta$, and $\tau$, one simultaneously seeks large $\depsTarget$ and small $|J|$.

%removed (\emph{i.e.}~$\Delta J/J \rightarrow 0$)
\subsubsection{Regimes I and II}
In regimes (I) and (II), $\depsTarget$ can be made arbitrarily large by increasing $L$ or $E_0$, respectively.  In both regimes, $\Delta J/J$ decreases as the dots become more isolated, and eventually Eq.~\ref{eqDEpsConstraint} will be satisfied.  In regime (I), $|J(\epsilon)|$ also decreases as the dots become more isolated, so that at large enough $L$ or $E_0$, Eq.~\ref{eqMagJConstraint} (or \ref{eqNoiseConstraintNOOPb}) will be satisfied.  In regime (II), $J$ takes a (usually non-zero) value dependent on $E_0$ and $B$ that can be made to satisfy Eq.~\ref{eqMagJConstraint}.  Thus, in regimes (I) and (II) one can theoretically satisfy Eqs.~\ref{eqMagJConstraint} and \ref{eqDEpsConstraint}, or Eq.~\ref{eqNoiseConstraintNOOPb} in the case of a no-op gate, by forming two isolated 1-electron dots or a single 2-electron dot.

%\begin{figure}[h]
%\begin{center}
%\includegraphics[width=1.7in,angle=270]{finalGoodCurve2.ps}
%\caption{Exchange energy $J$ vs.~bias $\epsilon$ where regions exist at low- and high  $\epsilon$ such that $\depsTarget \ge 0.1\meV$ for gate time $\tau=1\ns$ and error threshold $\Pth=10^{-4}$. \label{figGoodCurve2}}
%\end{center}
%\end{figure}

\begin{figure}[h]
\begin{center}
\includegraphics[width=2in,angle=270]{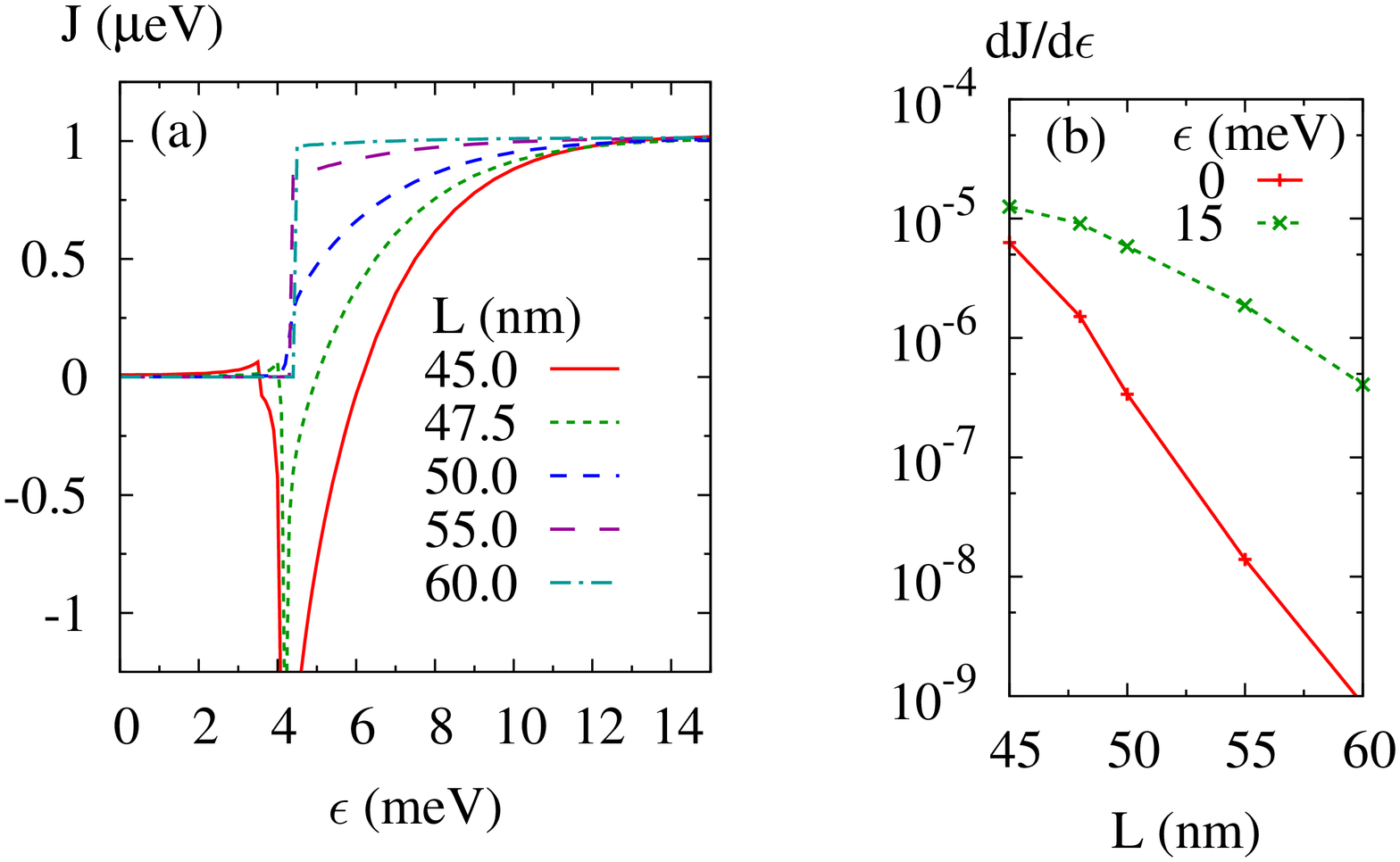}\vspace{-0.5cm}
\includegraphics[width=2in,angle=270]{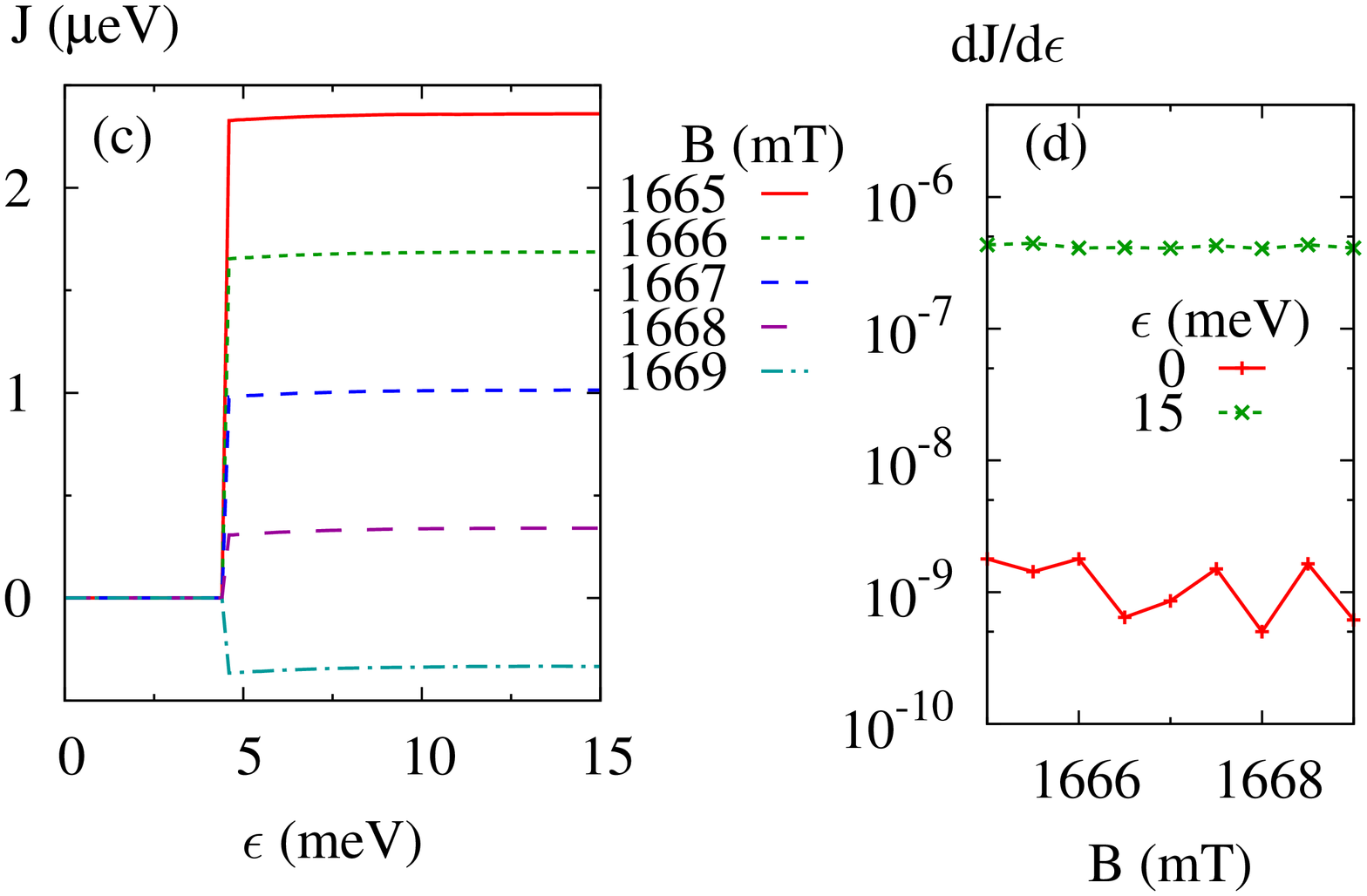}\vspace{-0.5cm}
\includegraphics[width=2in,angle=270]{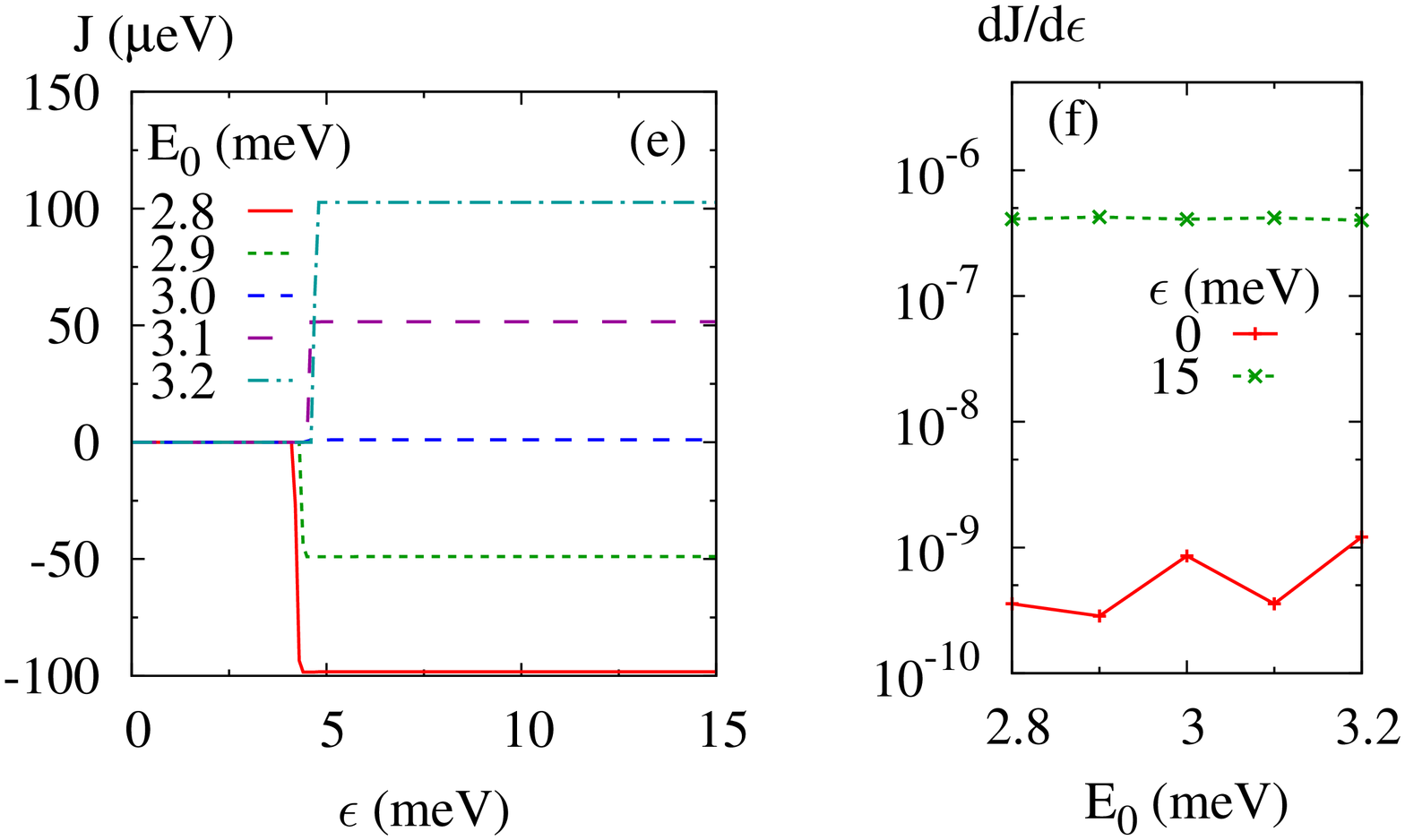}\vspace{-0.5cm}
\caption{Trends in the flat regions (regimes (I) and (II) in text) of exchange energy vs.~bias curves around the point $E_0=3\meV$, $L=60\nm$, and $B=1667\mT$. Frames (a), (c), and (e) show the dependence on $L$ (with $B=1667\mT$ and $E_0=3\meV$), $B$ (with $L=60$, and $E_0=3\meV$), and $E_0$ (with $L=60\nm$ and $B=1667\mT$), respectively. Frames (b), (d) and (f) show the derivative $dJ/d\epsilon$ of the curve at their left for $\epsilon=0$ and $15\meV$. \label{figPlateauTrends}}
\end{center}
\end{figure}

%HERE
Consider the system given by $E_0=3\meV$, $L=60\nm$, and $B=1667\mT$.  These parameters have been tuned within the range of physically reasonable values to result in an exchange curve with two relatively flat regions at low- and high-$\epsilon$.  This curve is shown in Fig.~\ref{figPlateauTrends} along with additional curves generated by varying $E_0$, $L$, and $B$ around the point $E_0=3\meV$, $L=60\nm$, and $B=1667\mT$.  

At low-$\epsilon$ (regime (I)),  $|J|<10^{-5}\mueV$ over a $3\meV$ window around $\epsilon=0$. Using Eq.~\ref{eqNoiseConstraintNOOPb} we find that the total idle time $\tau_0$ can be up to $(65.8\,\mu s)\sqrt{\Pth}$ ($66\ns$ when $\Pth=10^{-6}$), which is typically the time of many gate operations.  On the high-$\epsilon$ flat (regime II), $J \approx 1\mueV$ and $dJ/d\epsilon \approx 5\times 10^{-7}$.  This allows rotation gate times of order $1\ns$, and from Eq.~\ref{eqDEpsConstraint} $\depsTarget = (2\,\mbox{eV})\sqrt{\Pth}$ ($2\meV$ for $\Pth=10^{-6}$, which practically is limited by the width of the flat region).  Although these results are only semi-quantitative, this example shows that regimes exist for the DQD system where no-op and rotation operations are compatible with current quantum error correction and make realistic demands on controlling electronics technology.  This statement assumes, however, that only $\epsilon$-noise is present (\emph{i.e.} parameters $L$, $B$, and $E_0$ do not fluctuate).

In actuality, $L$, $B$, and $E_0$ cannot be perfectly controlled, and the variation of the exchange energy due to their fluctuations must be considered.  In regime (I), when both the singlet and triplet are in the (1,1) charge sector, $J$ is sensitive to the tunneling between the dots, and in general affected by $L$, $E_0$, and $B$.  Because the exchange energy is suppressed with increasing tunnel barrier, however, $J$ and its variation over a given $L$-, $E_0$-, or $B$-interval can be made arbitrarily small by choosing sufficiently large $L$, $E_0$ and/or $B$.  This is favorable for realizing a robust no-op.  In the high-$\epsilon$ regime (II), both electrons are almost completely confined to a single dot, and $J$ is strongly dependent on $E_0$ and $B$.  The inter-dot spacing, on the other hand, has a relatively small effect on $J$ that diminishes as $\epsilon$ increases.  In general, similar qualitative behavior is obtained by increasing $E_0$ or increasing $B$ (due to their common confining effect on the electrons).  This gives some freedom in selecting a dot size ($E_0$), and involves inherent trade-offs.  For instance, in large dots smaller magnetic fields can accomplish the same effects, but larger dots are also more susceptible to disorder effects (\emph{e.g.}~phonon induced spin-orbit coupling\cite{HansonSpinOrbit_2007}). %TODO: check/expand spin orbit stmtp
%, and are more difficult to control in the few-electron occupancy regime. %MAYBE KEEP (Malcolm?)

The dependence of $J$ on $L$, $B$, and $E_0$ in regimes (I) and (II) can be seen in Fig.~\ref{figPlateauTrends}a, c, and e, respectively.  Additionally, Fig.~\ref{figPlateauTrends}b shows that the derivatives in these regimes are sensitive to $L$ and therefore, even though changes in $L$ do not affect the value of $J$ on the upper flat, it must be sufficiently controlled that $dJ/d\epsilon$ remains within an acceptable range.  Overall, to utilize the $\epsilon$-noise robustness of regime (II) requires an ability to hold $L$, $B$ and $E_0$ fixed to the extent that Eq.~\ref{eqNoiseConstraint1} is satisfied.  Variations in $L$ are least problematic, since keeping the dots sufficiently isolated will ensure that $dJ/d\epsilon$ is small.  The typically strong linear dependence of $J$ on $B$ and $E_0$, however, could not be avoided in the parameter ranges we studied.

%Finally, we note that for weak enough tunelling between the dots (small $L$) we see the local minimum of regime (III) appear.

%Maybe add somewhere above?
%Note that the qualitative behavior resulting from changes in $B$ and $E_0$ is similar is due to the confining effect of magnetic field.

%OLD
%  Several general features of the $L$, $B$, and $E_0$ dependence shown by Fig.~\ref{figPlateauTrends} should be noted.  First, the value of $J$ in regime (II) is sensitive to changes in $B$ and $E_0$, and thus both must be controlled so to the extent that Eq.~\ref{eqNoiseConstraint1} is satisfied.  In regime (I), even though $J$ is sensitive to all three parameters, this can be overcome by sufficiently isolating the dots (see discussion above??).    Secondly, the derivatives in regimes (I) and (II) are sensitive to $L$ and therefore, even though changes in $L$ do not affect the value of $J$ on the upper flat, it must be sufficently controlled that $dJ/d\epsilon$ remains within an acceptable range.  
%As $B$ or $E_0$ is varied, the exchange energy at $\epsilon=0$ and $dJ/d\epsilon$ at $\epsilon=15\meV$ remain of the same order, so that the flats continue to be robust to changes in $\epsilon$.  However, unlike when varying $L$, the value pof $J$ at $\epsilon=15\meV$ has a strong linear dependence on $E_0$ and $B$, making the qubit operation very sensitive to noise in thsese quantities.

In summary, if $L$, $B$, and $E_0$ can be held fixed precisely enough (\emph{i.e.}~to satisfy Eq.~\ref{eqNoiseConstraint1}), then there exist regimes of the DQD system which realize a robust no-op and rotation operation by varying only the inter-dot bias $\epsilon$.  These robust regimes meet the requirements for current quantum error correction architectures and the control of $\epsilon$ falls within current the capabilities of state-of-the-art electronics.

%TODO: maybe add to caption??
%  .  Since all these curves are relatively flat around $\epsilon=0$ and $\epsilon=15\meV$, the derivative $dJ/d\epsilon$ can be used to calculate $\Delta J = (dJ/d\epsilon)(\Delta\epsilon)$.  The right-hand frames in Fig.~\ref{figPlateauTrends}  plot these derivatives separately as functions of $L$, $B$, and $E_0$.  

%For example, assuming a $\Pth$ of $10^{-4}$ and a gate time $1\ns$, it is possible to simultaneously achieve $\depsTarget \ge 0.1\meV$ for a no-op gate in low-$\epsilon$ regime (I) and a rotation gate in high-$\epsilon$ regime (II).  This involves tuning $B$ and $E_0$ so that a double-occupied dot has $J$ of order$\mueV$, and Fig.~\ref{figGoodCurve2} gives an example using realistic parameters.  Here it is possible to move between idle and rotation operating points by varying $\epsilon$ only. %where $E_0=3\meV$ and $B=1150\mT$.

We note, however, that a large (0,2) singlet-triplet splitting is necessary for loading and measurement so that the singlet can be selected with high probability.  Thus, if the high-$\epsilon$ regime is used for quantum operations, there must be a method of temporarily increasing the (0,2) exchange splitting during initialization and measurement.

\begin{figure}[h]
\begin{center}
\includegraphics[width=2in,angle=270]{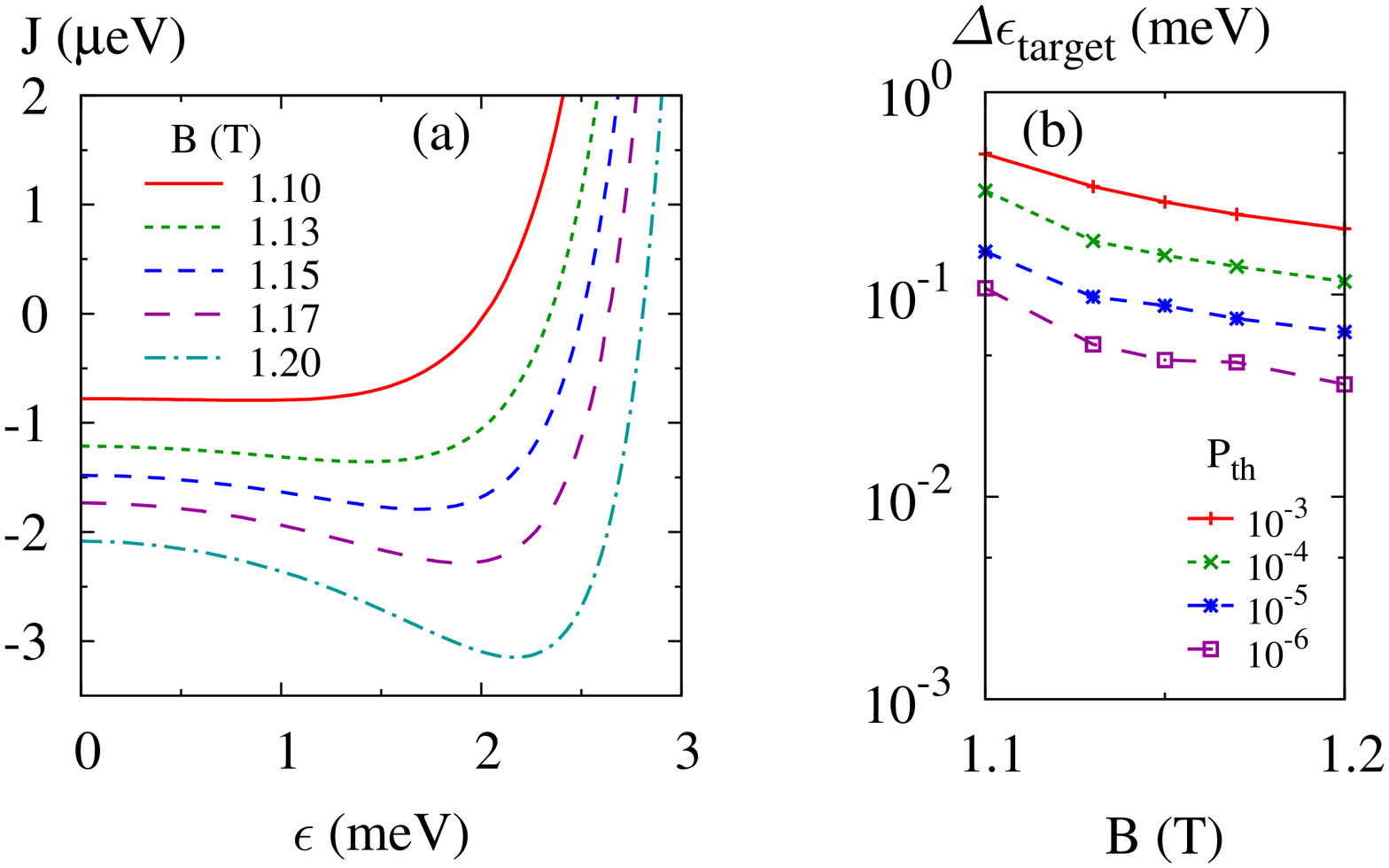}\vspace{-0.5cm}
\includegraphics[width=2in,angle=270]{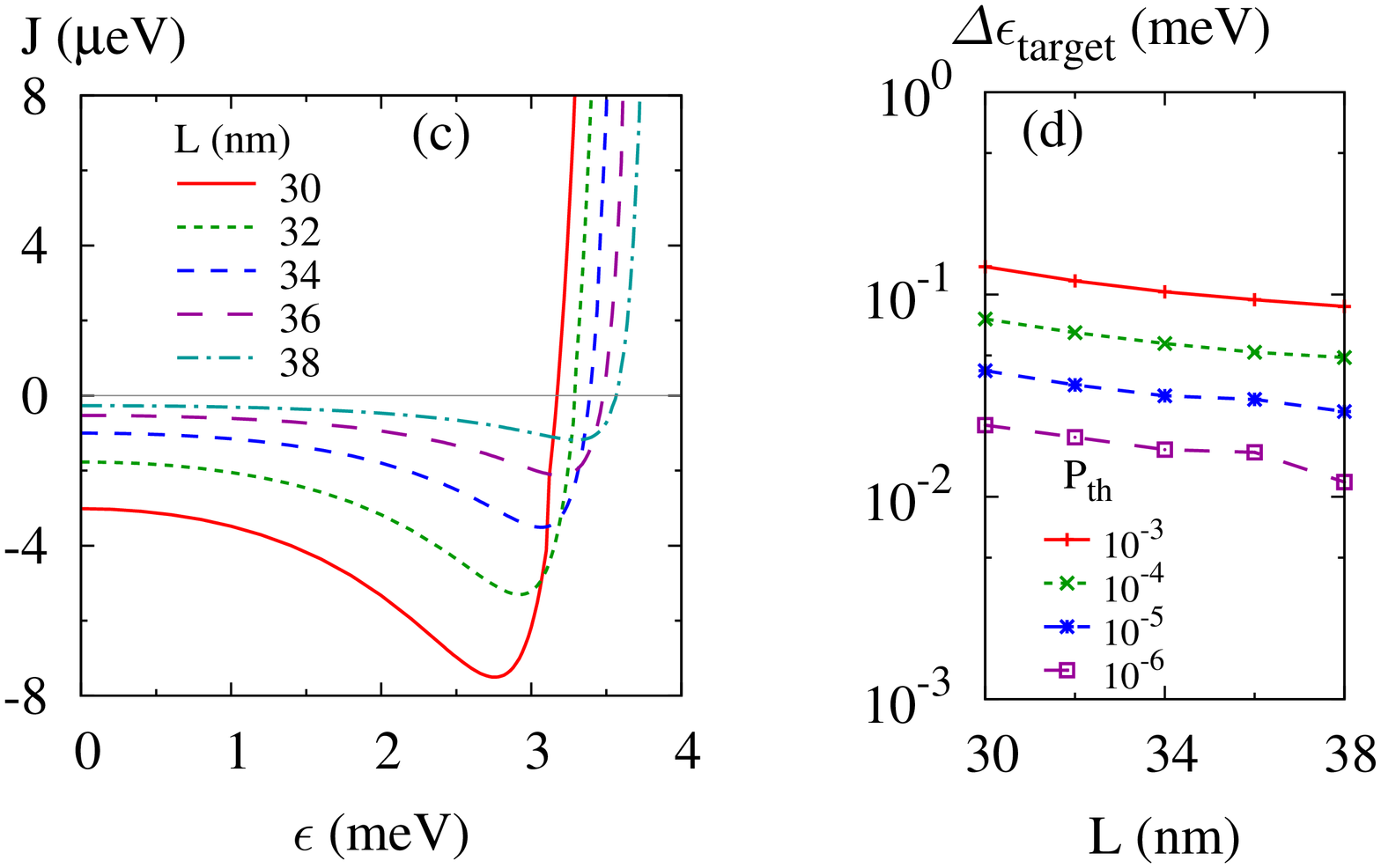}\vspace{-0.5cm}
\includegraphics[width=2in,angle=270]{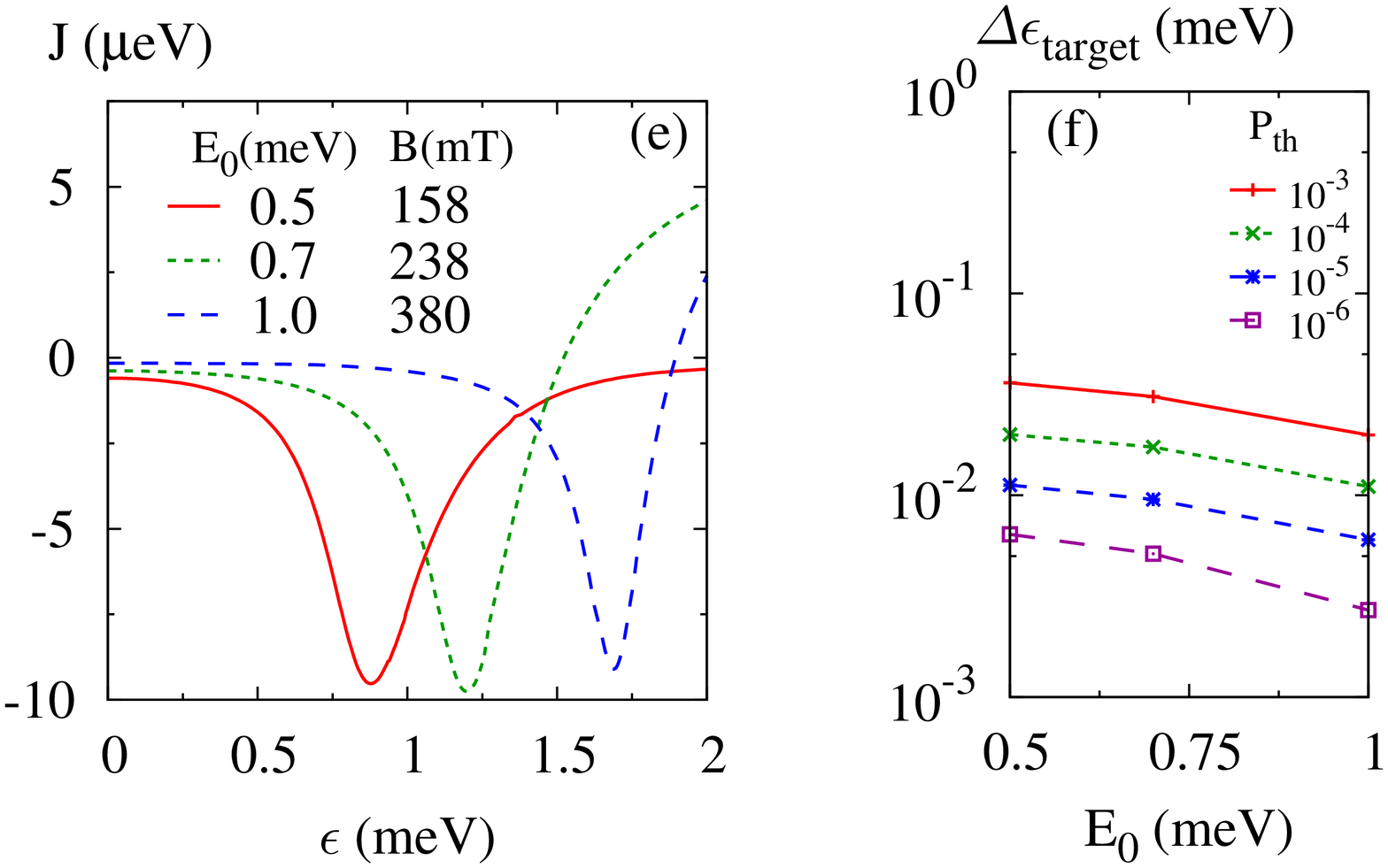}\vspace{-0.5cm}
\caption{Trends in relative curvature around the local minimum in exchange energy as a function of DQD bias $\epsilon$ (regime II in text).  Frames (a) and (b) show the dependence on $B$ (with $L=30$, and $E_0=3\meV$) and $L$ (with $B=1300\mT$ and $E_0=3\meV$), respectively.  Frames (c) and (d) plot the ratio $\Delta J/J$ as a function of $B$ and $L$, respectively, for the plots in frames (a) and (c).  It is seen that decreasing $B$ or decreasing $L$ correlates with smaller $\Delta J/J$, and this better robustness to $\epsilon$ variations.  In frame (e), curves with $L=60\nm$ and similar values at their minima, but different $E_0$, are shown, and it is seen that decreasing $E_0$ (and $B$) lead to greater $\epsilon$-robustness.\label{figDipTrends}}
\end{center}
\end{figure}

\subsubsection{Regime III}
In regime (III), $dJ/d\epsilon=0$ exactly at the minimum, and it more informative to study the behavior of $\depsTarget$ at the minimum, which is a measure of the curvature of $J(\epsilon)$ relative to its magnitude.  For a given $\Pth$, $\depsTarget$ can be increased by decreasing $B$, $L$, or $E_0$.  Decreasing $B$ gives less energetic advantage to the triplet state (relative to the singlet) and results in the minimum becoming less sharp as well as occurring at smaller $|J|$, as shown in Fig.~\ref{figDipTrends}a.  Overall, $\depsTarget$ increases as seen in Fig.~\ref{figDipTrends}b.  Either decreasing $L$ or $E_0$ increases the overlap between (1,1)- and (0,2)-states. Even though this pushes the minimum to larger $|J|$ because the (1,1) states have greater (negative) exchange energy, $\depsTarget$ increases because the (1,1)-(0,2) transitions of the singlet and triplet occur more gradually, and are farther separated in $\epsilon$.  %check this?
This dependence on $L$ is shown in Fig.~\ref{figDipTrends}c and d, and the dependence on $E_0$ in Figs.~\ref{figDipTrends}e and f. Note that in Fig.~\ref{figDipTrends} we compensate an increase in $E_0$ by reducing $B$ to obtain local minima which occur at similar values of $J$.  Even without this compensation, lower values of $E_0$ give larger $\depsTarget$ for fixed $\Pth$.   Because the singlet and triplet states have mixed (1,1) and (0,2) character in regime (III), variations in $L$, $E_0$, and $B$ are particularly effective at changing the charge distribution of the singlet and triplet, and thereby the exchange energy.  The sensitivity of $J$ to variation in $L$ is greater in this region than in either the low- or high-$\epsilon$ region, and the sensitivity to $E_0$ and $B$ lies between that of the low- and high-$\epsilon$ regions.

%Figure with all ``good curves'' grouped together
%\begin{figure}[h]
%\begin{center}
%\includegraphics[width=4.5in,angle=270]{finalGoodCurves.ps}
%\caption{Exchange energy curves which show potential operating points for a DQD qubit.  At points A through D in frames (a) and (b), the exchange energy's dependence on $\epsilon$ is compatible with a $1\ns$ gate time and $10^{-4}$ error threshold.  Points A and C are in the low-$\epsilon$ regime (I), point B is in regime (II), and point D is in the high-$\epsilon$ regime (III) (see text for regime descriptions).  Frame (c) shows a curve which has operating points in all three regimes, but requires a gate time of $0.1\ns$ and control of $\epsilon$ on $0.01\meV$ scales to be compatible with a $10^{-4}$ error threshold.\label{figGoodCurves}}
%\end{center}
%\end{figure}

\begin{figure}[h]
\begin{center}
\includegraphics[width=1.7in,angle=270]{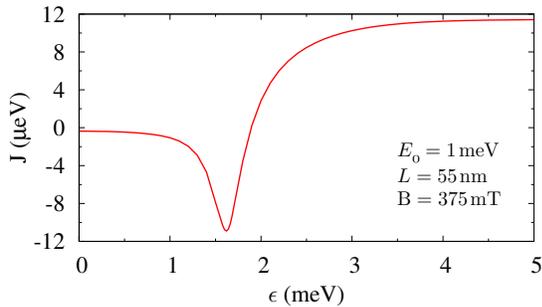}
\caption{Exchange energy $J$ vs.~bias $\epsilon$ with three potential operating points lying at $\epsilon \approx 0$, $1.6$, and $5\meV$.  To be compatible with an error threshold of $10^{-4}$, however, would require a gate time of order $0.1\ns$ and control of $\epsilon$ to a few hundredths of a micro-electronvolt. \label{figGoodCurve3}}
\end{center}
\end{figure}

%TODO: add subsubsection{XXX} here??
Ideally, all three regimes could be used as robust operating points simultaneously with the low-$\epsilon$ region serving as an idle point.  Multiple rotation speeds and negative rotation are potentially desirable for DCG.  Achieving this goal is challenging for several reasons.  The first is that broadening the regime-(III) minimum by coupling the dots more strongly is correlated with the exchange energy at $\epsilon=0$ increasing in magnitude (see Fig.~\ref{figDipTrends}).  This sets up a competition between a robust idle state in regime (I) requiring small $|J|$, and a robust operating point in regime (III), which requires a broad minimum.  Indeed, we find that for $\Pth=10^{-4}$ and $\tau=1\ns$ it is impossible to satisfy the constraints for an idle gate in regime (I) and a rotation gate in regime (III) at the same time.  Secondly, achieving a given value of $|J|$ at large-$\epsilon$ requires tuning either the dot size (via $E_0$) or magnetic field.  But since the magnetic field must be tuned to give a usable minimum in regime (III), tuning the dot size is necessary to simultaneously achieve a usable regime (II).  For $|J|$ of order $\mueV$ these constraints can lead to very large dots ($d$ greater than $200\nm$).  Figure \ref{figGoodCurve3} shows an example of a curve containing operation points in all three regimes (and with regime I an idle point).  Utilizing subsequent error correction, however, requires very fast gating of $0.1\ns$ and high detuning accuracy of $0.02\meV$, which represents a significant technological challenge.  In the end, we find that while the physics of the DQD system allows at least three robust operating points reachable by only changing the inter-dot bias, utilizing all three will require either finer electronics control or more efficient quantum error correction algorithms.

\section{Discussion\label{secDiscussion}}
It may not be necessary, however, to have three (or more) operating points separated only by changes in the inter-dot bias.  One could envision implementing two rotation speeds in a DQD qubit by varying another parameter, such as $E_0$ or $L$, as well as $\epsilon$, and to perform this alternate variation while $\epsilon=0$ and the qubit is in a robust no-op state. Or perhaps only one rotation speed will be necessary to begin with.  Our results indicate that a robust no-op should be accessible using current control and error correction technology, and that a robust rotation operation is also feasible as long as the shape of and spacing between the dots can be controlled with high precision.

Though dynamics are not studied in this work, it is important to realize that utilizing a (0,2)-flat region (regime II) requires the inter-dot bias to be quickly changed so that relatively little time is spent in the region of the $J$ curve between the low- and high-$\epsilon$ flats.  The speed at which this bias change occurs is limited by the gap to higher energy singlet and unpolarized triplet states as dictated by the adiabatic theorem.  We have considered such restrictions, and find that the gap to excited levels remains large enough that the qubit can be moved adiabatically between regimes I and II with the vast majority of the gate time spent in the noise robust regimes.  This does, however, set a bound on how weakly the dots can be coupled, since the gap to excited states decreases with the inter-dot coupling.

The architecture of the DQD can be used to mitigate the effects of the exchange energy's sensitivity to $L$ and $E_0$.  To the extent that the actual DQD potential remains a double-parabolic well, there will be a mapping from sets of gate voltages to the parameters $\epsilon$, $L$, and $E_0$.  Variations in these parameters is thus determined by their dependence on the gate voltages which vary to perform a qubit operation.  In this work, we have identified regions of ($\epsilon$,$L$,$E_0$)-space which are favorable for suppressing charge noise because they are flat, or nearly flat, along at least the $\epsilon$-direction.  By modifying the architecture of a DQD device, one can hope to map the pathways in gate-voltage space that perform qubit operations onto pathways in ($\epsilon$,$L$,$E_0$)-space that begin and end along flat regions.  In the present work we specifically focus on robustness to $\epsilon$ variations, and the ideal architecture would allow gate voltages to change $\epsilon$ while keeping $E_0$ and $L$ fixed.  For example, to keep $E_0$ fixed for the right dot, changes in inter-dot bias might be controlled exclusively by varying the voltages of gates around the left dot.  Such architecture engineering was first proposed by Friesen et al.,\cite{FriesenDigital_2002} where by moving the electrons along parallel channels instead of directly toward or away from each other, the inter-dot separation $L$ varies only quadratically in the gate voltages at a finite-$J$ operating point, instead of linearly.  In that work, however, since the DQD is in a regime where $J$ depended exponentially on $L$, the architecture serves only to reduce the sensitivity of $L$ to the gate voltages, not to map the pathway onto a flat curve in ($\epsilon$,$L$,$E_0$) parameter space.

\section{Conclusion\label{secConclusion}} 
In summary, configuration interaction calculations on a singlet-triplet DQD qubit using GaAs material parameters have been carried out. Three regimes have been identified in which the exchange energy is relatively insensitive to changes in the inter-dot bias $\epsilon$.  The CI method is necessary to both qualitatively and quantitatively calculate the dependence of $J$ on critical parameters such as detuning $\epsilon$, dot energy $E_0$, and dot separation $L$, compared to previous more approximate schemes such as HL or HM.  In particular, the CI method is found invaluable for calculations of critical regions such as when the dots are strongly coupled or when a single dot is doubly-occupied.  Namely, it captures the regime in which both the singlet and triplet transition into the (0,2) charge sector.  

By tuning only the inter-dot bias it is possible to travel between two or possibly three (with advances in electronics technology) of these robust regimes, which is desired for dynamically corrected gates and suggests how they might be implemented.  These types of calculations are needed to provide guidance regarding accuracy requirements for $\epsilon$, $E_0$, and $L$ given a QEC threshold.  We note the adverse effects caused by the sensitivity to certain parameters may be avoided by clever design of the qubit control electronics and architecture.

We would like to thank Sankar Das Sarma, Mike Stopa, and Wayne Witzel for many helpful discussions during the preparation of this manuscript.  This work was supported by the Laboratory Directed Research and Development program at Sandia National Laboratories. Sandia National Laboratories is a multi-program laboratory operated by Sandia Corporation, a wholly owned subsidiary of Lockheed Martin company, for the U.S. Department of Energy’s National Nuclear Security Administration under contract DE-AC04-94AL85000.

\vspace{-0.5cm}
\bibliography{noiseOptimal}

\end{document}